\documentclass{nature} 
\usepackage[latin1]{inputenc}
\usepackage[english]{babel}
\usepackage{enumerate}
\usepackage{color}

\usepackage{dcolumn}
\usepackage{bm}
\usepackage{mathptmx}
\usepackage{pifont}
\usepackage[latin1]{inputenc}
\usepackage{cancel}
\usepackage{amssymb}
\usepackage{amsmath}
\usepackage{upgreek}
\usepackage{xcolor}
\usepackage{graphicx}
\usepackage[normalem]{ulem}
\newcommand\bluesout{\bgroup\markoverwith{\textcolor{blue}{\rule[0.5ex]{2pt}{1.1pt}}}\ULon}
\makeatletter
\let\saved@includegraphics\includegraphics
\AtBeginDocument{\let\includegraphics\saved@includegraphics}
\renewenvironment*{figure}{\@float{figure}}{\end@float}
\makeatother
\title{The interplay of magnetic order with the electronic scattering and crystal-field effects in a metallic ferromagnet}

\author{Payel Shee,$^1$ Tanaya Halder,$^{1}$ Chia-Jung Yang,$^2$ Nainish Tickoo,$^1$ Ratiranjan Samal,$^{1}$  Ruta Kulkarni,$^3$ Shishir K. Pandey,$^{4,5}$ Vikas Kashid,$^{6,7}$ Ashis K. Nandy,$^{1}$ Arumugam Thamizhavel,$^{3}$ Anamitra Mukherjee,$^{1}$ and Shovon Pal$^1$}

\begin{document}
\maketitle
\begin{affiliations}
\item School of Physical Sciences, National Institute of Science Education and Research, An OCC of HBNI, Jatni, 752 050 Odisha, India
\item Department of Materials, ETH Zurich, 8093 Zurich, Switzerland
\item Department of Condensed Matter Physics and Materials Science, Tata Institute of Fundamental Research, 400 005 Mumbai, India
\item Department of General Sciences (Physics), Birla Institute of Technology and Science, Pilani, Dubai Campus, Dubai International Academic City, 345 055 Dubai, United Arab Emirates
\item Department of Physics, Birla Institute of Technology and Science, Pilani, Hyderabad Campus, 500 078 Telangana, India
\item Department of Physics, Savitribai Phule Pune University, 411 007 Pune, India
\item MIE-SPPU Institute of Higher Education, Doha, Qatar
\end{affiliations}

\begin{flushleft}
(Updated: \today)
\end{flushleft}

\begin{abstract}
The interplay between magnetic order, charge dynamics, and crystal field excitations underpins the emergent ground states of rare-earth intermetallics. Using time-domain terahertz spectroscopy, we probe this coupling in \text{PrSi}, a metallic ferromagnet. The optical response exhibits pronounced Drude-Smith behavior over a broad temperature range, indicating persistent carrier scattering. A classical Kondo-lattice model (CKLM) attributes this non-Drude conductivity to scattering of itinerant electrons by localized magnetic moments, persisting down to temperatures well below the magnetic ordering scale. At lower temperatures, beyond the scope of  CKLM, our experiment reveals that the response is dominated by crystal field excitations, with sharp transitions at 0.6\,THz and 1.54\,THz. The mode at 1.54\,THz shows a dynamic correlation with the onset of ferromagnetic order, marking the onset of a crystal-field-governed low temperature regime.
\end{abstract}

\maketitle
\section{Introduction}
Rare-earth intermetallic compounds offer a diverse platform to explore strong electronic correlations such as heavy-fermions~\cite{Coleman2006,Wirth2016}, topological insulators~\cite{Moore2010,Muniz2023}, quantum spin liquids~\cite{Broholm2020,Bairwa2025} and high-temperature superconductors~\cite{Maekawa1978}. In this class of materials, unpaired electrons in the $f$-orbitals of the rare-earth ions create localized magnetic moments that contribute to the long-range magnetic order in the  material. Here, the direct exchange interaction between the rare-earth ions is almost negligible because the overlap between the electronic shell of the neighboring ions is usually lower~\cite{Kirchmayr1978,Jensen1991}. The magnetic order is thus predominantly mediated by an indirect exchange interaction -- the Ruderman-Kittel-Kasuya-Yosida~(RKKY) interaction~\cite{Ruderman1954,Kasuya1956a,Yosida1957}. The ordering temperature (for example, the Curie temperature in ferromagnets) is dictated by the magnitude of the exchange energy scale~\cite{Savchenkov2023}. The spin-orbit coupling in addition plays a crucial role in the determination of the Fermi surface alongside the optical and transport properties. The interplay of itinerant carriers with the local moments has a long-standing history of uncovering phenomena such as Kondo screening and competing magnetic orders~\cite{Doniach1977}. It is well-known that in turn magnetic order or its lack has a significant bearing on carrier lifetime due to the local-moment-mediated scattering and can renormalize electron-electron interactions~\cite{Burch2008}. Evidently such correlated systems provide an extended opportunity to explore the interplay between their magnetic properties, carrier scattering rates and the optical properties. 

In the context of optical properties, both real ($\sigma_{\rm r}$) and imaginary ($\sigma_{\rm im}$) parts of the complex-valued THz conductivity carries information on the materials' electronic response to external light fields~\cite{Basov2011,Yang2023}. For example, electrons in a homogeneous medium, mostly follow the Drude law where the real part of the THz conductivity in the low frequency regime shows a predominantly higher value~\cite{Shimakawa2016}. However, for materials where the electrons see an inhomogeneous environment, such as confining structures or localization effects, the conductivity deviates from the regular Drude behavior. In most cases, when the dimension of the confinement becomes comparable to the carrier mean free path, non-Drude behavior sets in~\cite{Cocker2017}. In such scenarios, the Drude-Smith (DS) model has been a popular choice to account for the experimentally observed non-Drude-like THz conductivity in a wide range of materials, such as liquid metals~\cite{Smith1968,Smith2001,Jean2008}, nano-structured materials~\cite{Richter2010,Yang2013}, disordered crystals~\cite{Walther2007,Iwamoto2013,Wang2016} and molecular networks~\cite{Cooke2012,Jin2014,Puthukkudi2024}. The DS model is in fact an extension of the regular Drude model with the introduction of a `localization parameter' that quantifies the fraction of original velocity retained by the electrons after each scattering, or equivalently, the degree of backscattering in the system. In magnetic materials, Coulomb-interaction-mediated electronic scatterings and the exchange-interaction-driven magnetism are strongly correlated to each other. In systems with large magnetic moments allowing these to be treated classically, the relative orientation of the spins, in fact, influences the conductivity of a ferromagnetic metal. Above the Curie temperature, when the spins are randomly oriented, more scattering is introduced in the system, causing the system's conductivity to deviate from the ideal Drude behavior. Below Curie temperature the spins align, which should ideally reduce the scattering rate. In reality, however, the emergence of magnetic domains could affect the overall transport as well as the optical properties. It is known that the exchange interactions between the itinerant and the localized electrons often give rise to anomalous conductivity in ferromagnetic materials~\cite{Kasuya1956,Yang2003,Bonetti2016}. In addition, the presence of strong spin-orbit coupling can also lead to non-Drude transport, which has been demonstrated in ferromagnetic iron~\cite{Singh1975}.

While on one hand the magnetic ordering interferes with the Coulomb interaction in governing the transport characteristics of magnetic systems, the crystal electric field (CEF) on the other hand can potentially influence the magnetic ground state and anisotropy in these materials that concomitantly modifies the optical conductivity. Due to the periodic arrangement of the ligands/ions in the crystal structure, the CEF arises in the system, which breaks the degeneracy of $f$-orbitals in a rare-earth system. Depending on the total angular momentum ($J$) of the rare-earth ion, the CEF splits the ground state into ($2J+1$) sub-levels, where the spacing between the levels lies in the meV range. By breaking the degeneracy of the ground state, the CEF partially quenches the orbital contribution of the magnetic moment, resulting in a suppression of the total magnetic moment at temperatures close to the energy scales of the CEF splittings~\cite{Trammell1963}. The CEF thus becomes a pivotal factor in deciding the magnetic ground state of the rare-earth based systems. The interplay between the RKKY interaction and the CEF energy scales gives rise to several fascinating phenomena, namely, the exotic magnetic behavior in rare-earth based triangular-lattice quantum spin liquid systems~\cite{Scheie2020,Bordelon2019}, spin reorientation phase transition in rare-earth orthoferrites~\cite{Vovk2025} and the renormalized energy-scales of the CEF satellite states at quantum critical point in rare-earth based Kondo systems~\cite{Pal2019,Yang2020,Shee2024}. There are, however, limited work on the study of how the underlying CEF environment interferes the magnetic ordering in a rare-earth intermetallic compound having a ferromagnetic ground state.

In this article, we present the influence of magnetic ordering on the THz conductivity in a rare-earth based metallic ferromagnet exhibiting dominant crystal field environment at lower temperatures. For our investigations, we chose a Pr-based intermetallic compound, PrSi, which is metallic at all temperatures. PrSi crystallizes in an orthorhombic FeB-type crystal structure, as shown in Fig.~\ref{fig1}a, with the space group $Pnma$. Here, Pr is a rare-earth element, having a partially-filled 4$f$ orbital, which creates the magnetic moment in the material. Due to the Russel-Saunders coupling scheme, the central ion Pr$^{3+}$ gets a total angular momentum of $J = 4$, i.e, this ion posses a $(2J + 1) = 9$-fold degeneracy in its ground state. In presence of CEF, the ground state degeneracy splits into 9 singlets with an overall spacing of 284\,K~\cite{Das2014}. The ground and the first excited states are, however, separated by only 9\,K which is essentially responsible for the magnetic ordering in the material~\cite{Snyman2012}. This material undergoes a paramagnetic to ferromagnetic phase transition below $T_{\rm C}=52$\,K~\cite{Pinguet2003,Wang2014}. From earlier reports, we found that the [010]-axis is the easy magnetization axis of the material. The magnetic behavior in the other two axes are relatively complex. In particular, below 10\,K, the in-plane magnetization shows an anomalous behavior due to the presence of the CEF transition in the system. Through theoretical modeling of the magnetic susceptibility data, the CEF transitions in PrSi are predicted to lie in the THz range, specifically in the range $0.18-5.9$\,THz~\cite{Das2014}, see Fig.~\ref{fig1}b. Using THz time-domain spectroscopy, we observed an overall predominance of non-Drude behavior in the THz conductivity on either side of the phase transition. The non-Drude behavior supports the existence of carrier backscatterings in the system on either side of the phase transition. This behavior is phenomenologically explained using the classical Kondo lattice model (CKLM) up to a temperature of $T_{\rm CK}$ well below the $T_{\rm C}$ of the material. Since the phenomenological CKLM is used to model the magnetic scattering effects, it misses out on the CEF effects. In the experiment we continue to observe the non-Drude behavior till the lowest temperature measured. We argue that for temperatures lower than $T_{\rm CK}$, the carrier scattering introduced by the CEF environment most likely takes over leading to the observed non-Drude behavior. Upon a careful scrutiny of the conductivity spectra, we find evidences of CEF states that was theoretically predicted earlier~\cite{Das2014}. From the spectral weight analysis, we find that the relative occupation of the third CEF state corroborates the $T_{\rm C}$ of the material. Our observations thus highlight the underlying coupling between the CEF environment and the magnetic ordering in the material.

\section{Results and discussion}
\begin{figure}
	\centering
	\includegraphics[width=\linewidth]{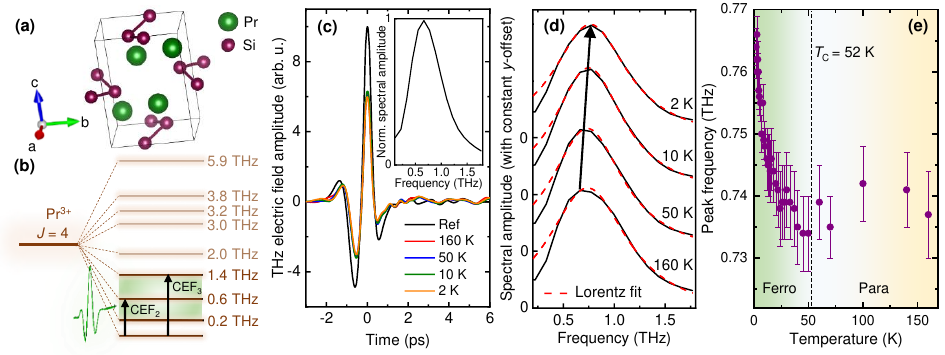}
	\caption{{\bf (a)} A schematic of the orthorhombic crystal structure of PrSi. {\bf (b)} A schematic showing the splitting of the $J = 4$ level of the central ion Pr$^{3+}$ into $2J+1 = 9$ sub-levels of distinct energy values, all in the THz range. The incident THz pulse interacts with the lowest four CEF multiplets. {\bf (c)} THz transients reflected from the reference Pt-mirror and the PrSi sample at  different temperatures. {\bf (Inset)} The incident THz spectra. {\bf (d)} The spectral amplitude corresponding to the sample at the corresponding temperatures shown in (c), obtained from the Fourier transform of the time transients. The zero-level corresponding to each spectra is marked. The red-dashed lines refer to the Lorentzian fitting of the respective spectra. The arrow indicates the peak shift as we lower the temperature. {\bf (e)} The temperature-dependent peak frequency corresponding to (d), which shows a blue shift as we enter in the ferromagnetic phase of the material. Here, the error bars are the standard errors from the modeling of the spectra. The vertical dashed line indicates the Curie temperature, $T_{\rm C} = 52$\,K.}
	\label{fig1}
\end{figure}

Temperature-dependent THz reflection measurements were performed to investigate the nature of the low-energy optical conductivity as PrSi goes through the paramagnetic to ferromagnetic phase transition. The exemplary THz time transients reflected from the PrSi single-crystal for a few temperatures along with the reference Pt-mirror are shown in Fig.~\ref{fig1}c. The inset in Fig.~\ref{fig1}c shows the typical THz spectrum that is incident on the sample. In addition to an overall absorption, we can observe the subtle indications of additional oscillatory features in the tail of the reflected THz pulses. The spectral response at different temperatures is obtained by taking a fast Fourier transform of the time transients, some of which at few distinct temperature points are shown in Fig.~\ref{fig1}d. We find that the spectral peak shows a continuous blue shift as the sample enters the ferromagnetic phase, i.e., below $T_{\rm C} = 52$\,K as shown in Fig.~\ref{fig1}e. We quantify this shift to be around 30\,GHz by using a Lorentzian function to fit the temperature-dependent reflection spectra. We speculate that the blue shift is associated with the enhanced RKKY-mediated exchange energy as material enters deeper into the ferromagnetic phase. Such exchange-mediated spectral blue shifts have been reported earlier in metallic~\cite{Pal2019,Shee2024} and semiconductor~\cite{Bossini2020} systems with antiferromagnetic ground state or in Mott systems~\cite{Hafez2021}.

\begin{figure}
	\centering
	\includegraphics[width=0.6\columnwidth]{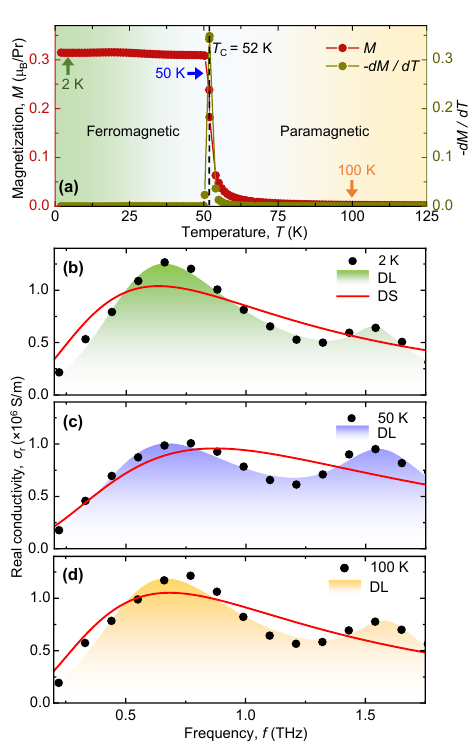}
	\caption{{\bf (a)} Temperature-dependent magnetization and its derivative, showing the $T_{\rm C}$ of the material at 52\,K. The real part of the THz conductivity as a function of frequency for {\bf (b)} 2\,K, {\bf (c)} 50\,K and {\bf (d)} 100\,K, respectively. The solid-red lines represent the Drude-Smith (DS) model. The colored areas represent the fitting using the double-Lorentz (DL) oscillator model.}
	\label{fig2}
\end{figure}

The $T_{\rm C}$ of 52\,K~\cite{Das2014, Snyman2012} was verified for the current sample by performing the magnetization measurements using a Quantum Design SQUID magnetometer. Figure~\ref{fig2}a, shows the temperature-dependent magnetization along the easy axis (i.e., [010] direction) in presence of an external out-of-plane magnetic field of 0.05\,T. The magnetization shows a sharp increase as the system evolves from paramagnetic phase (yellow shaded region) to the ferromagnetic phase (green shaded region). The Curie temperature quantified by taking a temperature derivative of the magnetization (i.e., $dM/dT$) as shown in Fig.~\ref{fig2}a, which shows a peak at 52\,K, marking the Curie temperature for our single-crystal PrSi. To understand the impact of magnetic ordering on the electron-electron interaction from the temperature-dependent reflected THz responses, we extract the real ($\sigma_{\rm r}$) and imaginary ($\sigma_{\rm im}$) parts of the THz conductivity. This is obtained using Fresnel's equation for the $p$-polarized light, as per the geometry of our experiments. Further elaborate details on the analysis procedure can be found elsewhere~\cite{Shee2024}.

PrSi by nature is metallic across the phase transition, indicating that one would expect the conductivity to follow Drude-like behavior. This is, however, in striking contrast to what we have observed in our experiments. We find that the conductivity is predominantly non-Drude-like. Deviations from the Drude behavior have been previously reported in several other metallic systems, such as heavy-fermions~\cite{Shee2024}, high-T$_{\rm c}$ superconductors~\cite{Schlesinger1990} and magnetic metals~\cite{Bonetti2016}. Figures~\ref{fig2}b,~\ref{fig2}c, and~\ref{fig2}d, show, respectively, the real part of the THz conductivity at 2\,K, 50\,K and 100\,K, corresponding to three different points in the PrSi phase diagram, namely, the ferromagnetic side, close to the transition point and the paramagnetic side. We use two different approaches to model our observation. In the first approach, we use the Drude-Smith (DS) model (as indicated by the red-solid lines in Figs.~\ref{fig2}b-d), motivated from the fact that the system hosts strong electronic interactions between the localized 4$f$-electrons of Pr and the itinerant conduction electrons. In the second approach, we use a double-Lorentz (DL) model (shown as colored areas in Figs.~\ref{fig2}b-d) to precisely account for the observed conductivity peaks, i.e., at around 0.6\,THz and 1.54\,THz. The possibility of magnons in this frequency range can be safely ignored as discussed in Ref.~\cite{Das2014}. We have also verified the absence of phonons in the measured frequency range using density-functional theory calculations (see the phonon calculations in Sec.~S1 of the Supplementary Information). We, thus, associate the observed peaks to the CEF states that has been theoretically proposed earlier~\cite{Das2014} to lie in the same frequency range.

\begin{figure}
	\centering
	\includegraphics[width=0.55\linewidth]{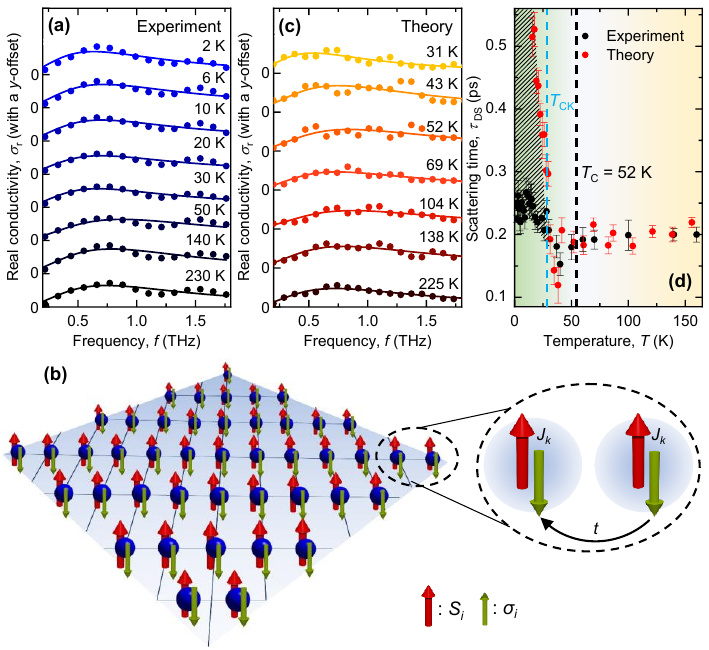}
	\caption{{\bf (a)} Temperature-dependent real part of the THz conductivity obtained from our experiments. The solid lines represent the fitting using Drude-Smith (DS) model. {\bf (b)} A schematic representation of the CKLM, where the localized 4$f$-moments (red arrows, $S_i$) are coupled to the itinerant electrons (green arrows, $\sigma_i$) at every lattice site. {\bf (Inset)} The localized and itinerant electrons are coupled with a coupling strength $J_k$ and $t$ being the nearest-neighbor hopping parameter. {\bf (c)} Temperature-dependent real part of the THz conductivity obtained from our theoretical CKLM. The solid lines represents fitting using Drude-Smith model. {\bf (d)} The scattering time $\tau_{\rm DS}$ obtained by fitting the Drude-Smith models corresponding to the experimental and theoretical conductivity spectra. The yellow and green-shaded regions show the paramagnetic and the ferromagnetic phases, respectively. The vertical black-dashed line marks the $T_{\rm C}$ of the material. The vertical cyan-dashed line marks the temperature below which the conductivity obtained from CKLM deviates from the experimental observations. Here, the error bars represent the standard errors from the DS-modeling of the experimental and theoretical conductivity data.}
	\label{fig3}
\end{figure}

We first scrutinize the real part of the THz conductivity using the DS model, where the complex conductivity, $\tilde \sigma_{\rm DS}(\omega) = \sigma_{\rm r}(\omega) + i\sigma_{\rm im}(\omega)$ is given by,
\begin{equation}\label{eq1}
\begin{split}
    \tilde \sigma_{\rm DS}(\omega)
    =\frac{\omega_{\rm p, DS}^2\epsilon_0\tau_{\rm DS}}{1-i\omega\tau_{\rm DS}}\left[{1+\frac{c}{1-i\omega\tau_{\rm DS}}}\right].    
\end{split}
\end{equation}
Here, $\omega_{\rm p, DS}$ (= 2$\pi f_{\rm p, DS}$) denotes the plasma frequency, $\tau_{\rm DS}$  is the scattering time, $\epsilon_0$ is the vacuum permittivity, and $c$ is the localization parameter, ranging from $-1$ to 0. A value of \( c = -1 \) corresponds to the maximum backscattering, where the electron reverses its direction completely after a scattering event. On the other hand, with \( c = 0 \) the DS model (i.e., Eq.~\ref{eq1}) reduces to the standard Drude model, with zero backscattering events. Figure~\ref{fig3}a shows the real part of the THz conductivity at several temperatures along with the DS model as solid-lines. It is quite evident the DS model remains dominant throughout the entire temperature range. Upon fitting the real part of Eq.~\ref{eq1} to the experimental data, we extracted the temperature-dependent scattering time ($\tau_{\rm DS}$) that carries the information on the nature of electronic interactions in the system, see the black dots in Fig.~\ref{fig3}d. We find that above $T_{\rm C}$ (i.e., in the paramagnetic phase), the scattering time is almost temperature-independent and is close to $\approx 0.2$\,ps. Below $T_{\rm C}$, however, $\tau_{\rm DS}$ increases and settles at a slightly higher value ($\approx 0.25$\,ps). Higher scattering time (alternately, reduced scattering rate) in the ferromagnetic phase clearly signals the existence of an intricate relationship between the electronic interactions and the magnetic ordering in the system. Such correlation (namely, exchange-induced electronic scattering) often leads to anomalies in DC resistivity~\cite{Peski-Tinbergen1963,Wu2009}. It is striking to note that the $\tau_{\rm DS}$ corroborates the $T_{\rm C}$ of the material to a very good extent. The temperature-dependent imaginary parts of the THz conductivity and the corresponding results obtained from the DS model fitting of the imaginary part of Eq.~\ref{eq1} are shown in Sec.~S2 of Supplementary Information. We note that we obtained identical values of scattering times from the fitting of the imaginary THz conductivity. We have, further, traced the temperature evolution of the localization parameter, $c$, obtained from the Drude-Smith model. This parameter gives a qualitative idea on the localization effects and hence a measure of the deviation from the ideal Drude transport. We find that the change in the localization parameter is maximum close to the phase transition, shown in Fig.~S3 of the Supplementary Information. This would indicate that close to the phase transition temperature, the material goes through a dramatic change in the electronic environment and an onset of the formation of confining structures such as domains, which potentially acts as source for an overall increase in the material's entropy and thereby sourcing the magnetocaloric effect -- a phenomenon that has been reported to occur in the same temperature range~\cite{Das2014}.

To gain a qualitative understanding on the correlation between the magnetic ordering and the electronic scattering rates, we formulated a classical Kondo lattice model (CKLM) at the phenomenological level. Our model considers the localized moments to be periodically arranged on a two-dimensional lattice, which are further coupled to the itinerant electrons. This is schematically shown in Fig.~\ref{fig3}b. In the rare earth materials, such local moments arise due to the Hund's coupling in the $f$-shells. Within the CKLM, the Hamiltonian $\mathcal{H}$ is given by
\begin{equation}\label{eq2}
    \mathcal{H}=-\sum_{<i,j>,\sigma}\left (t_{ij}c_{i\sigma}^{\dagger}c_{j\sigma}+h.c\right )+J_k\sum_{i}\textbf{S}_i\cdot\vec{\sigma_i}-\mu\sum_in_i.
\end{equation}
The first term of the Hamiltonian corresponds to the nearest-neighbor hopping that describes the motion of the conduction electrons in the lattice. The second term describes the coupling between the localized 4$f$-spins and the itinerant conduction electrons in the system. The last term is used to control the conduction electron density $n_i$, given by $n_i=\sum_{\sigma}c_{i\sigma}^{\dagger}c_{i\sigma}$ through the chemical potential $\mu$. Here, $t_{ij}=t=1$ is a uniform hopping amplitude, which sets the energy scale of the problem, $c_{i\sigma}$ and $c_{i\sigma}^{\dagger}$ are the fermion annihilation and creation operators, respectively, and $J_k$ is the local electron-spin coupling strength. In our model, we consider the local moments $\textbf{S}_i$ to be classical (mimicking the Pr $f$-orbital moments),  while $\vec{\sigma_i}$ is the spin of the conduction electron at site $i$, given by $\vec{\sigma_i}=\sum_{\alpha,\beta}c_{i\alpha}^{\dagger}\frac{\boldsymbol{\sigma}_{\alpha\beta}}{2}c_{i\beta}$, where the $\boldsymbol{\sigma}$ is composed of Pauli matrices. The model possesses full $SU(2)$ symmetry and has continuous $O(3)$ symmetry. It is well-known that such a model cannot have long-range magnetic order at finite temperatures. We employ an infinitesimal symmetry breaking magnetic field to stabilize magnetic order. We have checked the robustness of our results by varying the symmetry-breaking field strength. Using a reasonable value of $t=0.07\,e\rm V$, the value of scattering time, $\tau_{\rm DS}$ is matched with the experiments at higher temperatures. All temperature and frequency scales in the model are converted to the respective experimental units using this value of $t$. We note that the results reported here hold for $J_k/t\sim1.3$ to 1.5 on a square lattice where the ground state in the quantum limit is in the RKKY regime~\cite{Assaad1999,Assaad2019} and does not predict Kondo screening, consistent with earlier reports on PrSi~\cite{Das2014}.

Our model that underpins the coupling between the classical-spins and the itinerant fermions falls in the category of spin-fermion models~\cite{Mayr2005,Kumar2006,Mukherjee2013}, which is known~\cite{Patel2017,Mukherjee2014} to be solved using classical Monte-Carlo for annealing the classical degrees of freedom coupled with exact diagonalization for the fermion problem, allowing access to large lattice sizes~\cite{Kumar2006a,Mukherjee2015}. Using this exact diagonalization+Monte-Carlo (ED+MC) approach (as described in the methods section), we evaluate the frequency-resolved complex THz conductivity at several temperatures across the phase transition, the real part of which are shown in Fig.~\ref{fig3}c. We find that above $T=T_{\rm CK}$, the CKLM provides a very good agreement with our experimental observations, as shown by the red dots in Fig.~\ref{fig3}d. This clearly indicates that the low-energy transport is governed by the scattering between the localized 4$f$ spins and the itinerant electrons in our material system that ultimately results in the non-Drude nature of the transport. In the paramagnetic phase, when all the spins are randomly oriented, the electronic scattering within the system is high, i.e., the time between two scattering events (scattering time) is less. Below the Curie temperature, however, when the spins start to align, the electronic scattering decreases, leading to an increase of the scattering times. We note that below $T_{\rm CK}$ the scattering time obtained from the CKLM model deviates from the experimental ones. This is because in the CKLM, the scattering time diverges~\cite{Kasuya1956} once the classical spins order into a long-range uniform ferromagnet (for $T<T_{\rm CK}$). The model also allows us to calculate the inverse participation ratio (IPR) to qualitatively explore the localization properties of the itinerant electrons across the phase transition, see Sec.~S4 of the Supplementary Information for further details. We found that as we approach the $T_{\rm C}$ from the magnetic side, the enhanced disordering of the core spins increases the scattering, leading to an increased value of the IPR, evidently signifying an increased carrier localization, concomitant with the systematic suppression of the Drude weight as $\omega \to 0$~\cite{Kumar2005} for $T>T_{\rm C}$.

Our experiments however continues to exhibit non-Drude behavior. It is natural to expect the persistence of magnetic domains in the ferromagnetic phase in our experiments down to much lower temperature than in the CKLM, the dimensions of which lie in the same order of magnitude as the carrier mean free path~\cite{Sakarya2005}. We anticipate that this could also potentially introduce scattering centers in the magnetic phase at lower temperatures. Further, as discussed below, our experiments suggest another source of non-Drude behavior originating from the underlying crystal field environment, which is, however, beyond the scope of the simple CKLM.

\begin{figure}
	\centering
	\includegraphics[width=0.75\linewidth]{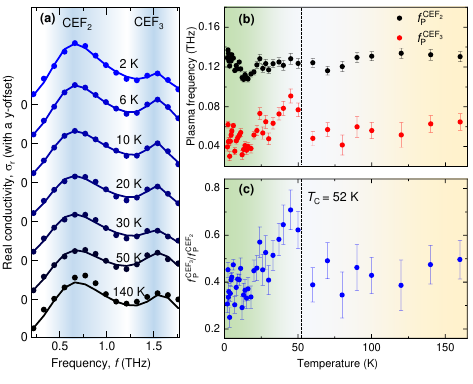}
	\caption{{\bf (a)} Real part of the THz conductivity at different temperatures. The solid lines represents fitting using the double-Lorentz (DL) model. {\bf (b)} The plasma frequency corresponding to both oscillators, plotted as a function of temperature. {\bf (c)} The ratio of the plasma frequencies in (b). The yellow- and green-shaded regions show the paramagnetic and the ferromagnetic phase, respectively, while the vertical dashed-line marks the Curie temperature of the system. Here, the error bars are the standard errors from the modeling of the experimental conductivity data.}
	\label{fig4}
\end{figure}

To substantiate the presence of crystal field environment, we choose a different approach where we take a closer look at the temperature-dependent spectral features in the real part of the THz conductivity. We find that the conductivity peaks at two frequencies, namely 0.6\,THz and 1.54\,THz (see Fig.~\ref{fig4}a). The peak positions are in good agreement with the reported second and third CEF transitions in PrSi~\cite{Das2014}, see Fig.~\ref{fig1}b. Since the first CEF state lies at 0.2\,THz, it is beyond our spectral resolution to probe it. We, thus, capture the next two CEF transitions at 0.6\,THz and 1.54\,THz. For our reference in the subsequent discussions, we name these transitions as CEF$_2$ and CEF$_3$. From the experimental real conductivity, shown in Fig.~\ref{fig4}a, we observe that the spectral amplitude of both these peaks display a distinct temperature dependence. We notice that the spectral amplitudes of the two peaks become almost comparable near $T_{\rm C}$. To quantify this behavior, we fit the experimental curves using double-Lorentz (DL) oscillator model, shown by the solid lines in Fig.~\ref{fig3}a that is given by,
\begin{equation}\label{eq:5}
    \sigma_{\rm DL}({\omega})= \sum_{i=1}^{2} \epsilon_0\omega_{pi, \rm DL}^2\frac{\omega\tau_{i,\rm DL}}{\omega+i\tau_{i,\rm DL}(\omega_{0i}^2-\omega^2)}.
\end{equation}
Here, $\omega_{pi, \rm DL} = 2\pi f_{pi, \rm DL} = \sqrt{n_ie^2/m^*\epsilon_0}$ is the plasma frequency. $\tau_{i,\rm DL}$ is the relaxation time, and $\omega_{0i}$ is peak frequency corresponding to the DL model (where $i =1, 2$). $n_i$ is the electronic density involved in the CEF transition, $e$ is the electronic charge, and $m^*$ is the effective mass of electrons. From our model fitting, we extracted the plasma frequency, which is a measure of the density of bound electrons per unit volume that contribute to the specific resonant CEF state absorption. The respective plasma frequencies are plotted in Fig.~\ref{fig4}b as a function of temperature. We find that the plasma frequency corresponding to the CEF$_2$ is always higher than the CEF$_3$, which implies a higher density of electrons involved in CEF$_2$ transition as compared to the CEF$_3$ transition. This is evident since CEF$_2$ is energetically lower than CEF$_3$. Surprisingly, close to the phase transition, the density of electrons involved in CEF$_3$ transition increases significantly, as plotted in Fig.~\ref{fig4}c. We have alternately verified our observations by calculating the frequency-integrated spectral weights using the Kubo's formula (see Sec.~S6 of Supplementary Information). On lowering the temperature below $T_{\rm C}$, the CEF$_3$ occupation relaxes back. Our results clearly highlights an intrinsic correlation between the CEF occupation and the onset of magnetic ordering. We note that while the CKLM works primarily for the high temperature phase, the Lorentz oscillator model works throughout the entire temperature range. The line-widths (i.e., $1/\tau_{i,\rm DL}$) of both the CEF peaks further substantiates the model validation, where we can clearly note an increase in its value with the onset of the magnetic order in the system, see Sec.~S7 of Supplementary Information. These low temperature observations also highlight the need of incorporating CEF in a Kondo lattice model treated with more sophisticated techniques such as Abrikosov fermion approach~\cite{Dong2021} to retain low temperature quantum fluctuations, providing an ample scope for further work along this direction. 

As PrSi goes from the paramagnetic (disordered) to the ferromagnetic (ordered) state, there is a change in the entropy associated with it, which leads to a change in the temperature of the material. The crystal electric field plays an important role in this process because it induces strong magnetic anisotropy in the system~\cite{Ranke2007,Tkac2015,Ranke2022}. The presence of the strong interplay between the crystal field and the magnetic ordering, as highlighted in our work through an enhanced occupation of third CEF state near $T_{\rm C}$ (shown in Fig.~\ref{fig4}c), strongly corroborates the previous observation of giant magnetocaloric effect in these materials~\cite{Das2014,Snyman2012}, an analogy of which is drawn in Fig.~S6b of the Supplementary Information. It is known that for rare-earth materials to have a high magnetocaloric property, the rare-earth should possess a high angular momentum with a high exchange interaction energy~\cite{Terada2023}, both of which are reported in PrSi~\cite{Das2014}.

\section{Conclusion}
In conclusion, our investigation on the temperature-dependent THz conductivity in a metallic ferromagnet, PrSi, provides a fresh outlook on the intrinsic interplay of crystal field environment on the magnetic ordering and the correlation between the magnetic $T_{\rm C}$ and the electronic scatterings. In particular, we observed a predominant non-Drude transport across the paramagnetic to ferromagnetic phase transition. This non-Drude response is associated with the carrier scattering events within the system, where the carrier scattering rate increases below $T_{\rm C}$. The behavior is phenomenologically modeled to a very good extent using a classical Kondo-lattice model, which highlights scattering interactions between the localized 4$f$-spins of Pr$^{3+}$ and the itinerant electrons. The model reproduces the experiments until a temperature that we mark as $T_{\rm CK}$, below which we argue that the underlying crystal field environment takes over. A more elaborate analysis of our conductivity data shows the presence of two CEF peaks that reveals a distinct temperature behavior, where the occupation of third CEF state peaks close to the phase transition. Our experiments clearly signals the presence of a dynamic correlation between the magnetic ordering and the crystal field environment. This correlation is, however, not unique to PrSi and can be extended to other rare-earth silicides $R$Si ($R=$ Nd, Ho, Dy, and Tb)~\cite{Nguyen1977,Nirmala2008,Schobinger2011,Kumar2024,Tanusilp2022}, Pr$_5$Si$_3$~\cite{Souptel2004,Tian2010}, $R$Ni ($R=$ Pr, Gd, Tb, Dy, Ho, and Er) compounds~\cite{Tripathy2005,Kumar2008,Rajivgandhi2017,Savchenkov2023} and $R$Al$_2$ compounds~\cite{Ranke2007}. Given the wide applicability, we anticipate that our results open up an ample scope for both theoretical and experimental developments on identifying such intrinsic correlations between the spin and charge degrees of freedom in a broader class of strongly interacting systems.

\clearpage

\clearpage
\section*{Experimental and Numerical Methods}
{\bf Sample preparation.}
Single crystal of PrSi was synthesized using the Czochralski pulling technique in a tetra-arc furnace (Technosearch Corporation, Japan), maintained under an argon atmosphere. Since PrSi melts congruently at 1657$^{\circ}$C, the high-purity starting elements, Pr (99.9\%) and Si (99.999\%) were taken in a stoichiometric 1:1 ratio. The mixture was melted multiple times in the tetra-arc furnace to ensure compositional uniformity. A tungsten rod was used as a seed crystal, which was pulled at a constant rate of 10\,mm/h. Samples from the same ingot were previously used for the investigation of electrical resistivity, magnetization, specific heat, and magneto-caloric effects~\cite{Das2014}. 

For our current investigations, we cut the sample surface oriented perpendicular to the crystallographic $b$-axis using a wire electric discharge machine. The sample surface is polished using colloidal silica, ensuring that any residual roughness is within the sub-micron range (i.e., $<<\lambda_{\rm THz}$). Given the metallic nature of PrSi across all temperatures, the experiments are carried out in the reflection geometry using linearly-polarized THz radiation with a spectral range of $0.2-2.5$\,THz, incident at an angle of 45$^{\circ}$. The electric field vector of the incident radiation lies within the crystallographic $ac$-plane. The sample is then mounted in a temperature-controlled helium reservoir cryostat for the temperature-dependent studies.

{\bf THz time-domain experiments.}
Single-cycle THz pulses are generated by optical rectification using a 0.5\,mm thick ZnTe crystal oriented along the (110)-plane, utilizing 90\% of the output from a Ti:Sapphire laser (central wavelength 800\,nm, pulse duration 120\,fs, repetition rate 1\,kHz, and a pulse energy 2\,mJ). The remaining 10\% of the beam is used as the gating pulse for free-space electro-optic detection of the reflected THz pulses. Both the reflected THz pulse and the gating beam are allowed to co-propagate through a second (110)-oriented ZnTe detection crystal. The THz-induced ellipticity in the gating beam is analyzed using a combination of a quarter-wave plate, a Wollaston prism, and a balanced photodiode detector. To suppress the Fabry-Perot resonances caused by the internal reflections within the 0.5-mm thick detection crystal, a 2-mm thick THz-inactive (100)-oriented ZnTe crystal is optically bonded to its rear surface, thereby extending the accessible time window. All measurements are performed in an inert nitrogen environment. A 15-nm Pt film grown on a quartz substrate is used as the reference and placed next to the sample within the cryostat to ensure identical experimental conditions.

{\bf Monte Carlo method.}
The classical Kondo lattice model involves both classical and quantum degrees of freedom that are coupled to each other. For this purpose, we use the ED+MC method to study the model at finite temperatures~\cite{Pradhan2007,Pradhan2009}. Our method follows a two-step process: $(i)$ update the classical spin at a site, $(ii)$ exactly diagonalize the full Hamiltonian for fixed classical background. We start by generating a random configuration of the spins on the entire lattice, say $\{\textbf{S}_i\}$. The $\textbf{S}_i$ has a fixed value and it can orient in any arbitrary direction in the 3-$D$ space. The Hamiltonian is then generated for this classical configuration, $\mathcal{H}(\{\textbf{S}_i\})$, and is diagonalized to find the energy, say $E_1$. We then propose an update at site, say $j$, and update the spin configuration on that particular site $\textbf{S}_j$ and re-diagonalize the Hamiltonian with that updated spin configuration to get an energy $E_2$. If the new energy with updated spin configuration is less than the old energy, $(E_1-E_2)>0$, then the update is accepted and the total spin configuration, $\{\textbf{S}_i\}$, is updated. If $(E_1-E_2)<0$, then the update is accepted or rejected based on the Boltzmann probability. At any particular temperature this process is repeated by sequentially visiting all the lattice sites. This constitutes a single system sweep. A thermalized regime is reached after a large number of system sweeps at a fixed temperature as discussed below. Once we reach the thermalized regime, we calculate the desired observables from the equilibrium configurations .

In our calculations, we start the simulation at a high temperature with a random spin configuration and then cool down the system in small steps of temperature so that the system does not get stuck in an intermediate metastable state. Here, the simulations are performed on an $8\times 8$ square lattice, using $2000+2000$ Monte-Carlo steps, where the first 2000 steps are used to thermalize the system and the later 2000 steps are used to calculate the observables. We calculate the observables after every 10$^{\rm th}$ step in the later part to avoid any autocorrelation.

For the characterization of the magnetic ordering in our system, we look at the magnetic structure factor $S(q)$, given by
\begin{equation*}
 S(q)=\frac{1}{N^2}\sum_{i,j}e^{i\textbf{q}\cdot(r_i-r_j)}\left<\textbf{S}_i\cdot\textbf{S}_j\right >,
\end{equation*}
where $\textbf{q}={0,0}$ is the wave-vector considered, as we are interested in the ferromagnetic order. $N$ is the dimension of the system. $\textbf{S}_i$ and $\textbf{S}_j$ are the spins at the $i-$th and $j-$th site, respectively. $r_i$ and $r_j$ give the position of the respective spins. The structure factor gives an estimation for the ordering temperature, see Fig.~S2 of the Supplementary Information. We further evaluate the optical conductivity of the system, using
\begin{equation*}
    \sigma(\omega)=\frac{\pi e^2}{N\hslash a_0}\sum_{\alpha\beta}\left(n_{\alpha}-n_{\beta}\right )\frac{|f_{\alpha\beta}|^2}{\epsilon_{\beta}-\epsilon_{\alpha}}\delta(\omega-(\epsilon_{\beta}-\epsilon_{\alpha})),
\end{equation*}
where $f_{\alpha\beta}$ are the matrix elements of the current operator, explicitly given by $f_{\alpha\beta}=\left <\psi_{\alpha}\right |J_x\left |\psi_{\beta}\right >$. $J_x$ being the current operator, given by
\begin{equation*}
    J_x=-ia_0t\sum_{i,\sigma}\left(c_{i,\sigma}^{\dagger}c_{i+a_0\hat{x}}-h.c.\right).
\end{equation*}    
Here, $\left|\psi_{\alpha}\right>$ and $\epsilon_{\alpha}$ are the single-particle eigenstates and eigenvalues, respectively. $n_{\alpha}=f(\mu-\epsilon_{\alpha})$ is the Fermi function and $a_0$ is the lattice parameter.

\begin{addendum}
\item[Author Contributions] All authors contributed to the discussion and interpretation of the experiment and to the completion of the manuscript. P.S. and C.J.Y. performed the experiments. P.S. and N.T. performed the data analysis. R.K. and A.T. provided the samples. R.S. performed the magnetization measurements. T.H., V.K., A.K.N. and A.M. developed the theoretical model, while T.H. and P.S. analyzed the theoretical results. S.K.P. and A.K.N. performed the phonon calculations. A.T. and S.P. conceived the project while S.P. supervised the experiments. P.S., T.H., A.M. and S.P. drafted the manuscript.

\item[Acknowledgements] P.S., A.K.N. and S.P. acknowledge the support from DAE through the projects Basic Research in Physical and Multidisciplinary Sciences via RIN4001 and New Frontiers in Earth, Atmospheric, Planetary, Stellar Physics, Material Sciences and Rare Event Searches via RNI4011. S.P. also acknowledges the start-up support from DAE through NISER and SERB through SERB-SRG via Project No.~SRG/2022/000290. T.H. and A.M. acknowledge the use of the NOETHER cluster at NISER. The authors thank A. K. Nayak for useful discussions and  assistance with the SQUID measurements.

\item[Competing Interests] The authors declare that they have no competing financial interests.

\item[Data Availability] {The datasets analyzed in the current study are attached. Any additional data are available from corresponding authors upon reasonable request.}

\item[Code Availability] {The codes associated with the Monte-Carlo method, supporting this study are available from the corresponding authors upon reasonable request.}

\item[Correspondence]{All correspondence should be addressed to S.P. (email: shovon.pal@niser.ac.in) or A.M. (email: anamitra@niser.ac.in) or P.S. (email: payel.shee@niser.ac.in)} 
\end{addendum}


\clearpage
\begin{thebibliography}{1}
\bibitem{Coleman2006}
P. Coleman, \emph{Heavy Fermions: Electrons at the edge of magnetism} (John Wiley \& Sons, New York, 2007).

\bibitem{Wirth2016}
S. Wirth, and F. Steglich, Exploring heavy fermions from macroscopic to microscopic length scales, Nat. Rev. Mater. {\bf 1}, 1 (2016). 

\bibitem{Moore2010}
J. E. Moore, The birth of topological insulators, Nature {\bf 464}, 194 (2010). 

\bibitem{Muniz2023}
B. M. Cano, Y. Ferreiros, P. A. Pantale{\'o}n, J. Dai, M. Tallarida, A. I. Figueroa, V. Marinova, K. Garc{\'\i}a-D{\'\i}ez, A. Mugarza, S. O. Valenzuela, R. Miranda, J. Camarero, F. Guinea, J. A. Silva-Guill{\'\i}n, and M. A. Valbuena, Experimental demonstration of a magnetically induced warping transition in a topological insulator mediated by rare-earth surface dopants, Nano Lett. {\bf 23}, 6249 (2023).

\bibitem{Broholm2020}
C. Broholm, R. J. Cava, S. A. Kivelson, D. G. Nocera, M. R. Norman, and T. Senthil, Quantum spin liquids, Science {\bf 367}, eaay0668 (2020).

\bibitem{Bairwa2025}
D. Bairwa, A. Bandyopadhyay, D. Adroja, G. B. G. Stenning, H. Luetkens, T. J. Hicken, J. A. Krieger, G. Cibin, M. Rotter, S. Rayaprol, P. D. Babu, and S. Elizabeth, Quantum spin liquid ground state in the rare-earth triangular antiferromagnet SmTa$_7$O$_{19}$, Phys. Rev. B {\bf 111}, 104413 (2025).

\bibitem{Maekawa1978}
S. Maekawa, and M. Tachiki, Superconducting phase transitions in rare-earth compounds, Phys. Rev. B {\bf 18}, 4688 (1978).

\bibitem{Kirchmayr1978}
H. R. Kirchmayr, and C. A. Poldy, Magnetism in rare earth-3\textit{d} intermetallics, J. Magn. Magn. Mater. {\bf 8}, 1 (1978). 

\bibitem{Jensen1991} 
J. Jensen, and A. R. Mackintosh, Rare earth magnetism: structures and excitations, Oxford University Press (1991).

\bibitem{Ruderman1954}
M. A. Ruderman, and C. Kittel, Indirect exchange coupling of nuclear magnetic moments by conduction electrons, Phys. Rev. {\bf 96}, 99 (1954).

\bibitem{Kasuya1956a}
T. Kasuya, A theory of metallic ferro- and antiferromagnetism on Zener's Model, Prog. Theor. Phys. {\bf 16}, 45 (1956).

\bibitem{Yosida1957}
K. Yosida, Magnetic properties of $\text{Cu-Mn}$ alloys, Phys. Rev. {\bf 106}, 893 (1957).

\bibitem{Savchenkov2023}
P. S. Savchenkov, and P. A. Alekseev, Uncommon magnetism in rare-earth intermetallic compounds with strong electronic correlations, Crystals {\bf 13}, 1238 (2023). 

\bibitem{Doniach1977}
S. Doniach, The Kondo lattice and weak antiferromagnetism, Physica B+C {\bf 91}, 231 (1977).

\bibitem{Burch2008}
K. S. Burch, D. D. Awschalom, and D. N. Basov, Optical properties of III-Mn-V ferromagnetic semiconductors, J. Magn. Magn. Mater. {\bf 320}, 3207 (2008).

\bibitem{Basov2011}
D. N. Basov, R. D. Averitt, D. van der Marel, M. Dressel, and K. Haule, Electrodynamics of correlated electron materials, Rev. Mod. Phys. {\bf 83}, 471 (2011).

\bibitem{Yang2023}
C.-J. Yang, J. Li, M. Fiebig, and S. Pal, Terahertz control of many-body dynamics in quantum materials, Nat. Rev. Mater. {\bf 8}, 518 (2023).

\bibitem{Shimakawa2016}
K. Shimakawa, and S. Kasap, Dynamics of carrier transport in nanoscale materials: Origin of Non-Drude Behavior in the terahertz frequency range, Appl. Sci. {\bf 6}, 50 (2016).

\bibitem{Cocker2017}
T. L. Cocker, D. Baillie, M. Buruma, L. V. Titova, R. D. Sydora, F. Marsiglio, and F. A. Hegmann, Microscopic origin of the Drude-Smith model, Phys. Rev. B {\bf 96}, 205439 (2017).

\bibitem{Smith1968}
N. V. Smith, Drude theory and the optical properties of liquid mercury, Phys. Lett. A {\bf 26}, 126 (1968).

\bibitem{Smith2001}
N. V. Smith, Classical generalization of the Drude formula for the optical conductivity, Phys. Rev. B {\bf 64}, 155106 (2001).

\bibitem{Jean2008}
J. Cl\'erouin, P. Noiret, V. N. Korobenko, and A. D. Rakhel, Direct measurements and ab initio simulations for expanded fluid aluminum in the metal-nonmetal transition range, Phys. Rev. B {\bf 78}, 224203 (2008).

\bibitem{Richter2010}
C. Richter and C. A. Schmuttenmaer, Exciton-like trap states limit electron mobility in TiO$_2$ nanotubes, Nat. Nanotechnol. {\bf 5}, 769 (2010).

\bibitem{Yang2013}
C.-S. Yang, C.-M. Chang, P.-H. Chen, P. Yu, C.-L. Pan, Broadband terahertz conductivity and optical transmission of indium-tin-oxide (ITO) nanomaterials, Opt. Express {\bf 21}, 16670 (2013).

\bibitem{Walther2007}
M. Walther, D. G. Cooke, C. Sherstan, M. Hajar, M. R. Freeman, and F. A. Hegmann, Terahertz conductivity of thin gold films at the metal-insulator percolation transition, Phys. Rev. B {\bf 76}, 125408 (2007).

\bibitem{Iwamoto2013}
K. Iwamoto, T. Mori, S. Kajitani, H. Matsumoto, N. Toyota, K. Suekuni, M. A. Avila, Y. Saiga, and T. Takabatake, Optical conductivity spectra of rattling phonons and charge carriers in the type-VIII clathrate Ba$_8$Ga$_{16}$Sn$_{30}$, Phys. Rev. B {\bf 88}, 104308 (2013).

\bibitem{Wang2016}
T. Wang, E. A. Romanova, N. Abdel-Moneim, D. Furniss, A. Loth, Z. Tang, A. Seddon, T. Benson, A. Lavrinenko, and P. U. Jepsen, Time-resolved terahertz spectroscopy of charge carrier dynamics in the chalcogenide glass As$_{30}$Se$_{30}$Te$_{40}$, Photo. Res. {\bf 4}, A22 (2016).

\bibitem{Cooke2012}
D. G. Cooke, F. C. Krebs, and P. U. Jepsen, Direct observation of sub-100 fs mobile charge generation in a polymer-fullerene film, Phys. Rev. Lett. {\bf 108}, 056603 (2012).

\bibitem{Jin2014}
Z. Jin, D. Gehrig, C. Dyer-Smith, E. J. Heilweil, F. Laquai, M. Bonn, and D. Turchinovich, Ultrafast terahertz photoconductivity of photovoltaic polymer-fullerene blends: A comparative study correlated with photovoltaic device performance, J. Phys. Chem. Lett. {\bf 5}, 3662 (2014).

\bibitem{Puthukkudi2024}
A. Puthukkudi, S. Nath, P. Shee, A. Dutta, C. V. Rajput, S. Bommakanti, J. Mohapatra, M. Samal, S. Anwar, S. Pal, and B. P. Biswal, Terahertz conductivity of free-standing 3D covalent organic framework membranes fabricated via triple-layer-dual interfacial approach, Adv. Mater. {\bf 36}, 2312960 (2024).

\bibitem{Kasuya1956}
T. Kasuya, Electrical resistance of ferromagnetic metals, Prog. Theor. Phys. {\bf 16}, 58 (1956).

\bibitem{Yang2003}
S.-R. E. Yang, J. Sinova, T. Jungwirth, Y. P. Shim, and A. H. MacDonald, Non-Drude optical conductivity of (III, Mn)V ferromagnetic semiconductors, Phys. Rev. B {\bf 67}, 045205 (2003).

\bibitem{Bonetti2016}
S. Bonetti, M. C. Hoffmann, M.-J. Sher, Z. Chen, S.-H. Yang, M. G. Samant, S. S. P. Parkin, and H. A. D\"{u}rr, THz driven ultrafast spin-lattice scattering in amorphous metallic ferromagnets, Phys. Rev. Lett. {\bf 117}, 087205 (2016).

\bibitem{Singh1975}
M. Singh, C. S. Wang, and J. Callaway, Spin-orbit coupling, Fermi surface and optical conductivity of ferromagnetic iron, Phys. Rev. B {\bf 11}, 287 (1975).

\bibitem{Trammell1963}
G. T. Trammell, Magnetic ordering properties of rare-earth ions in strong cubic crystal fields, Phys. Rev. {\bf 131}, 932 (1963).

\bibitem{Scheie2020}
A. Scheie, V. O. Garlea, L. D. Sanjeewa, J. Xing, and A. S. Sefat, Crystal-field Hamiltonian and anisotropy in KErSe$_2$ and CsErSe$_2$, Phys. Rev. B {\bf 101}, 144432 (2020).

\bibitem{Bordelon2019}
M. M. Bordelon, E. Kenney,  C. Liu, T. Hogan, L. Posthuma,  M. Kavand, Y. Lyu, M. Sherwin, N. P. Butch, C. Brown, M. J. Graf, L. Balents, and S. D.  Wilson, Field-tunable quantum disordered ground state in the triangular-lattice antiferromagnet NaYbO$_2$, Nat. Phys. {\bf 15}, 1058 (2019).

\bibitem{Vovk2025}
N. R. Vovk, E. V. Ezerskaya, and R. V. Mikhaylovskiy, Theory of terahertz-driven magnetic switching in rare-earth orthoferrites: The case of TmFeO$_3$, Phys. Rev. B {\bf 111}, 064411 (2025).

\bibitem{Pal2019}
S. Pal, C. Wetli, F. Zamani, O. Stockert, H. von L\"{o}hneysen, and M. Fiebig, Fermi volume evolution and crystal field excitations in heavy-fermion compounds probed by time-domain terahertz spectroscopy, Phys. Rev. Lett. {\bf 122}, 096401 (2019).

\bibitem{Yang2020}
C.-J. Yang, S. Pal, F. Zamani, K. Kliemt, C. Krellner, O. Stockert, H. von L\"{o}hneysen, J. Kroha, and M. Fiebig, Terahertz conductivity of heavy-fermion systems from time-resolved spectroscopy, Phys. Rev. Research {\bf 2}, 033296 (2020).

\bibitem{Shee2024}
P. Shee, C.-J Yang, S. K. Pandey, A. K. Nandy, R. Kulkarni, A. Thamizhavel, M. Fiebig, and S. Pal, Terahertz crystal electric field transitions in a Kondo-lattice antiferromagnet, Phys. Rev. B {\bf 109}, 075133 (2024).

\bibitem{Das2014}
P. K. Das, A. Bhattacharyya, R. Kulkarni, S. K. Dhar and A. Thamizhavel, Anisotropic magnetic properties and giant magnetocaloric effect of single-crystal PrSi, Phys. Rev. B {\bf 89}, 134418 (2014).

\bibitem{Snyman2012}
J. L. Snyman, and A. M. Strydom, Anomalous magnetic ground state in PrSi evidenced by the magnetocaloric effect, J. Appl. Phys. {\bf 111}, 07A943 (2012).

\bibitem{Pinguet2003}
N. Pinguet, F. Weitzer, K. Hiebl, J. C. Schuster, and H. No{\"e}l, Structural chemistry, magnetism and electrical properties of binary Pr-silicides, J. Alloys Compd. {\bf 348}, 1 (2003).

\bibitem{Wang2014}
L.-C. Wang, and B.-G. Shen, Magnetic properties and magnetocaloric effects of PrSi, Rare Met. {\bf 33}, 239 (2014).

\bibitem{Bossini2020}
D. Bossini, M. Terschanski, F. Mertens, G. Springholz, A. Bonanni, G. S. Uhrig, and M. Cinchetti, Exchange-mediated magnetic blue-shift of the band-gap energy in the antiferromagnetic semiconductor MnTe, New J. Phys. {\bf 22}, 083029 (2020).

\bibitem{Hafez2021}
M. Hafez-Torbati, D. Bossini, F. B. Anders, and G. S. Uhrig, Magnetic blue shift of Mott gaps enhanced by double exchange, Phys. Rev. Research {\bf 3}, 043232 (2021).

\bibitem{Schlesinger1990}
Z. Schlesinger, R. T. Collins, F. Holtzberg, C. Feild, and S. H. Blanton, U. Welp, G. W. Crabtree, Y. Fang, and J. Z. Liu, Superconducting energy gap and normal-state conductivity of a single domain YBa$_2$Cu$_3$O$_7$ crystal, Phys. Rev. Lett. {\bf 65}, 801 (1990).

\bibitem{Peski-Tinbergen1963}
T. Van Peski-Tinbergen, and A. J. Dekker, Spin-dependent scattering and resistivity of magnetic metals and alloys, Physica {\bf 29}, 917 (1963).

\bibitem{Wu2009}
D. Wu, N. Bari$\breve{s}$i$\acute{c}$, N. Drichko, S. Kaiser, A. Faridian, M. Dressel, S. Jiang, Z. Ren, L. J. Li, G. H. Cao, Z. A. Xu, H. S. Jeevan, and P. Gegenwart, Effects of magnetic ordering on dynamical conductivity: Optical investigations of EuFe$_2$As$_2$ single crystals, Phys. Rev. B {\bf 79}, 155103 (2009).

\bibitem{Assaad1999}
F. F. Assaad, Quantum Monte Carlo simulations of the half-filled two-dimensional Kondo lattice model, Phys. Rev. Lett. {\bf 83}, 796 (1999).

\bibitem{Assaad2019}
M. Raczkowski, and F. F. Assaad, Emergent coherent lattice behavior in Kondo nanosystems, Phys. Rev. Lett. {\bf 122}, 097203 (2019).

\bibitem{Mayr2005}
M. Mayr, G. Alvarez, C. $\c{\rm S}$en, and E. Dagotto, Phase fluctuations in strongly coupled $d$-wave superconductors, Phys. Rev. Lett. {\bf 94}, 217001 (2005).

\bibitem{Kumar2006}
S. Kumar, and P. Majumdar, Insulator-metal phase diagram of the optimally doped manganites from the disordered Holestein-double exchange model, Phys. Rev. Lett. {\bf 96}, 016602 (2006).

\bibitem{Mukherjee2013}
A. Mukherjee, W. S. Cole, P. Woodward, M. Randeria, and N. Trivedi, Theory of strain-controlled magnetotransport and stabilization of the ferromagnetic insulating phase in manganite thin films, Phys. Rev. Lett. {\bf 110}, 157201 (2013).

\bibitem{Patel2017}
N. D. Patel, A. Mukherjee, N. Kaushal, A. Moreo, and E. Dagotto, Non-Fermi liquid behaviour and continuously tunable resistivity exponents in the Anderson-Hubbard model at finite temperature, Phys. Rev. Lett. {\bf 119}, 086601 (2017).

\bibitem{Mukherjee2014}
A. Mukherjee, N. D. Patel, S. Dong, S. Johnson, A. Moreo, and E. Dagotto, Testing the Monte Carlo-mean field approximation in the one-band Hubbard model, Phys. Rev. B {\bf 90}, 205133 (2014).

\bibitem{Kumar2006a}
S. Kumar, and P. Majumdar, A travelling cluster approximation for lattice fermions strongly coupled to classical degrees of freedom, Eur. Phys. J. B {\bf 50}, 571 (2006).

\bibitem{Mukherjee2015}
A. Mukherjee, N. D. Patel, C. Bishop, E. Dagotto, Parallelized traveling cluster approximation to study numerically spin-fermion models on large lattices, Phys. Rev. E {\bf 91}, 063303 (2015).

\bibitem{Kumar2005}
S. Kumar, and P. Majumdar, Transport and localisation in the presence of strong structural and spin disorder, Eur. Phys. J. B {\bf 46}, 237 (2005).

\bibitem{Sakarya2005}
S. Sakarya, N. H. van Dijk, and E. Br{\"u}ck, Determination of the magnetic domain size in the ferromagnetic superconductor UGe$_2$ by three-dimensional neutron depolarization, Phys. Rev. B {\bf 71}, 174417 (2005).

\bibitem{Dong2021}
J.-J. Dong, D. Huang, Y.-F. Feng, Mutual information, quantum phase transition, and phase coherence in Kondo systems, Phys. Rev. B {\bf 104}, L081115 (2021).

\bibitem{Ranke2007}
P. J. von Ranke, N. A. de Oliveira, D. C. Garcia, V. S. R. de Sousa, V. A. de Souza, A. Magnus G. Carvalho, S. Gama, and M. S. Reis, Magnetocaloric effect due to spin reorientation in the crystalline electrical field: Theory applied to DyAl$_2$, Phys. Rev. B {\bf 75}, 184420 (2007).

\bibitem{Tkac2015}
V. Tk{\'a}\v{c}, A. Orend{\'a}\v{c}ov{\'a}, E. \v{C}i\v{z}m{\'a}r, M. Orend{\'a}\v{c}, A. Feher, and A. G. Anders, Giant reversible rotating cryomagnetocaloric effect in KEr(MoO$_4$)$_2$ induced by a crystal-field anisotropy, Phys. Rev. B. {\bf 92}, 024406 (2015).

\bibitem{Ranke2022}
P. J. von Ranke, and W. S. Torres, Theoretical investigation of crystalline electric field influence on the magnetocaloric effect in the cubic praseodymium system PrNi$_2$, Phys. Rev. B {\bf 105}, 085153 (2022).

\bibitem{Terada2023}
N. Terada, H. Mamiya, H. Saito, T. Nakajima, T. D. Yamamoto, K. Terashima, H. Takeya, O. Sakai, S. Itoh, Y. Takano, M. Hase, and H. Kitazawa, Crystal electric field level scheme leading to giant magnetocaloric effect for hydrogen liquefaction, Commun. Mater. {\bf 4}, 13 (2023).

\bibitem{Nguyen1977}
V. N. Nguyen, F. Tch\'eou, and J. Rossat-Mignod, Magnetic structures of PrSi and NdSi intermetallic alloys, Solid State Commun. {\bf 23}, 821 (1977).

\bibitem{Nirmala2008}
R. Nirmala, A. V. Morozkin, D. Buddhikot, and A. K. Nigam, Magnetocaloric effect in the binary intermetallic compound DySi, J. Magn. Magn. Mater. 320, 1184 (2008).

\bibitem{Schobinger2011}
P. Schobinger-Papamantellos, K. H. J. Buschow, and J. Rodr\'iguez-Carvajal, Magnetic phase diagrams of the CrB- and FeB-type HoSi compounds, J. Magn. Magn. Mater. 323, 2592 (2011).

\bibitem{Kumar2024}
A. Kumar, P. Singh, A. Doyle, D. L. Schlagel, and Y. Mudryk, Multiple magnetic interactions and large inverse magnetocaloric effect in TbSi and TbSi$_{0.6}$Ge$_{0.4}$, Phys. Rev. B {\bf 109}, 214410 (2024).

\bibitem{Tanusilp2022}
S. A. Tanusilp, K. Kurosaki, Rare-earth silicides: the promising candidates for the thermoelectric applications at near room temperature, Jpn. J. Appl. Phys. {\bf 62}, SD0802 (2022).

\bibitem{Souptel2004}
D. Souptel, A. Leithe-Jasper, W. L\"oser, W. Schnelle, H. Borrmann, and G. Behr, Floating zone growth and characterization of Pr$_5$Si$_3$ single crystals, J. Cryst. Growth {\bf 273}, 311 (2004).

\bibitem{Tian2010}
G. Tian, H. Du, Y. Zhang, Y. Xia, C. Wang, J. Han, S. Liu, and J. Yang, Large reversible magnetocaloric effect of light rare-earth intermetallic compound Pr$_5$Si$_3$, J. Alloys Compd. {\bf 496}, 517 (2010).

\bibitem{Tripathy2005}
S. K. Tripathy, K. G. Suresh, R. Nirmala, A. K. Nigam, and S. K. Malik, Magnetocaloric effect in the intermetallic compound DyNi, Solid State Commun. {\bf 134}, 323 (2005).

\bibitem{Kumar2008}
P. Kumar, K. G. Suresh, A. K. Nigam, and O. Gutfleisch, Large reversible magnetocaloric effect in RNi compounds, J. Phys. D: Appl. Phys. {\bf 41}, 245006 (2008).

\bibitem{Rajivgandhi2017}
R. Rajivgandhi, J. A. Chelvane, S. Quezado, S. K. Malik, and R. Nirmala, Effect of rapid quenching on the magnetism and magnetocaloric effect of equiatomic rare earth intermetallic compounds RNi (R = Gd, Tb and Ho), J. Magn. Magn. Mater. {\bf 433}, 169 (2017).


\bibitem{Pradhan2007}
K. Pradhan, A. Mukherjee, and P. Majumdar, Distinct effects of homogeneous weak disorder and dilute scatterers on phase competition in manganites, Phys. Rev. Lett. {\bf 99}, 147206 (2007).

\bibitem{Pradhan2009}
K. Pradhan, and P. Majumdar, Magnetic order beyond RKKY in the classical Kondo lattice, Europhys. Lett. {\bf 85}, 37007 (2009).

\end{thebibliography}
\end{document}


\maketitle
\begin{affiliations}
\item School of Physical Sciences, National Institute of Science Education and Research, An OCC of HBNI, Jatni, 752 050 Odisha, India
\item Department of Materials, ETH Zurich, 8093 Zurich, Switzerland
\item Department of Condensed Matter Physics and Materials Science, Tata Institute of Fundamental Research, 400 005 Mumbai, India
\item Department of General Sciences (Physics), Birla Institute of Technology and Science, Pilani, Dubai Campus, Dubai International Academic City, 345 055 Dubai, United Arab Emirates
\item Department of Physics, Birla Institute of Technology and Science, Pilani, Hyderabad Campus, 500 078 Telangana, India
\item Department of Physics, Savitribai Phule Pune University, 411 007 Pune, India
\item MIE-SPPU Institute of Higher Education, Doha, Qatar
\end{affiliations}

\begin{flushleft}
(Updated: \today)
\end{flushleft}

\begin{abstract}
This supplementary information contains further details on phonon modes in PrSi at THz frequencies, imaginary part of the THz conductivity, localization parameter, Inverse Participation Ratio (IPR), the magnetic structure factor, the frequency-integrated spectral weight of the real part of THz conductivity and the CEF line-widths. The contents are sectionized as:

\begin{description}
    \item S1. Phonon calculations 
    \item S2. Imaginary part of THz conductivity
    \item S3. Localization parameter
    \item s4. Inverse Participation Ratio (IPR)
    \item S5. The structure factor
    \item S6. Frequency-integrated spectral weight
    \item S7. CEF line-widths
\end{description}

\end{abstract}

\section{Phonon calculations}
Density-functional theory (DFT) calculations are performed using projector-augmented wave method \cite{PAW,PAWpotentials1} as implemented within the Vienna $Ab$-$initio$ Simulation Package (VASP)~\cite{Kresse}. The generalized-gradient approximation (GGA) with Perdew-Burke-Ernzerhof (PBE) functional form is used for the calculation of exchange-correlation energy~\cite{PBE}. For Pr and Si atoms, 4$f^3$5$s^2$5$p^6$ and 3$s^2$3$p^2$ are considered as valence configurations, respectively. The full lattice optimization of the orthorhombic primitive crystal structure of PrSi (SG: $Pnma$, \# 62) is done with energy and Hellmann-Feynman force convergence criteria of 10$^{-7}$\,eV and 10$^{-3}$\,eV/\AA{}, respectively. We used a Plane-Wave cutoff energy of 350\,eV and an 11 $\times$ 7 $\times$ 5 $\Gamma$-centered $k$-mesh for Brillouin zone sampling. A Gaussian smearing width of 0.05\,eV was considered in our calculations. We find a small deviation ($<$ 0.8\%) in lattice parameters during the optimization of the crystal structure. Phonon dispersion is then calculated using the finite differences method considering a 2 $\times$ 2 $\times$ 2 supercell. The post-processing of phonon calculations is done using Phonopy software~\cite{phonopy,phonopy2}. The Infrared (IR) active modes are analyzed using the SAM module~\cite{sam} of the Bilbao Crystallographic Server.

\begin{figure}[t!]
    \centering
	\includegraphics[width=0.7\columnwidth]{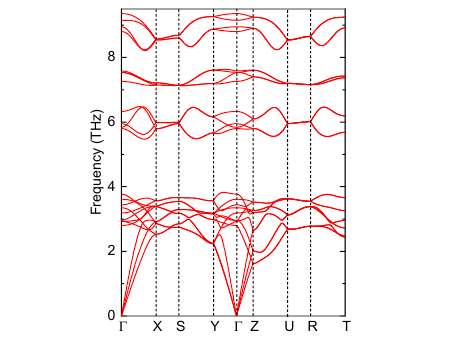}
    \vspace{-25pt}
	\caption{The calculated phonon dispersion using the finite differences method with a 2 $\times$ 2 $\times$ 2 supercell of PrSi.}
	\label{figS1}
\end{figure}

\begin{table}
 \centering
 \caption{Irreducible representations of IR-active phonon (Irreps) modes at the zone center are listed for PrSi corresponding to the 4$c$ Wyckoff positions.\\}
    \begin{tabular}{cc}
    \hline \hline
    Frequency & Irreps\\
    \hline
    2.80\,THz & $B_{3u}$\\ 
    2.94\,THz & $B_{1u}$\\
    5.74\,THz & $B_{3u}$\\  
    5.95\,THz & $B_{1u}$\\
    6.33\,THz & $B_{2u}$\\
    7.26\,THz & $B_{3u}$\\
    7.58\,THz & $B_{1u}$\\
    \hline \hline
    \end{tabular}
    \label{tab:ir}
\end{table}

The calculated phonon dispersion of PrSi is shown in Fig.~\ref{figS1}. Consistent with the experimental finding, in our calculations as well, we could not find any unstable phonon modes.  The IR active modes for PrSi can be found using the SAM module~\cite{sam} of the Bilbao Crystallographic Server. There are three IR-active modes allowed corresponding to the 4$c$ Wyckoff positions of Pr and Si, irreducible representations for which are $B_{1u}$, $B_{2u}$, and $B_{2u}$. The frequency of these modes is listed in Table~\ref{tab:ir} at the $\Gamma$-point.

\begin{figure}[t!]
    \centering
    \includegraphics[width=0.7\columnwidth]{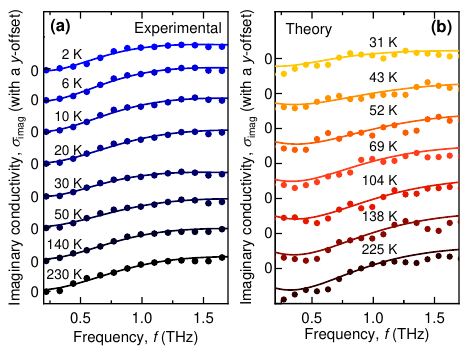}
    \vspace{-20pt}
	\caption{{\bf (a)} Imaginary part of THz conductivity obtained from our experiments. {\bf (b)} Imaginary part of THz conductivity obtained from the classical Kondo-lattice model. The solid lines represent fitted curves using the Drude-Smith model.}
	\label{figS2}
\end{figure}

\section{Imaginary part of THz conductivity} 
The temperature-dependent imaginary part of THz conductivity is plotted in Fig.~\ref{figS2}a. The solid lines depict the Drude-Smith model as per the imaginary part of Eq.1 of main manuscript.  Figure~\ref{figS2}b shows the imaginary part of THz conductivity as obtained from the CKLM via the Kramers-Kr\"onig transformation, supporting our experimental data above $T_{\rm CK}$. In addition, we have fitted the theoretically modeled data using Drude-Smith model, which shows close agreement with the experimental data. 

\section{Localization parameter}
\begin{figure}[t!]
    \centering
    \includegraphics[width=0.5\columnwidth]{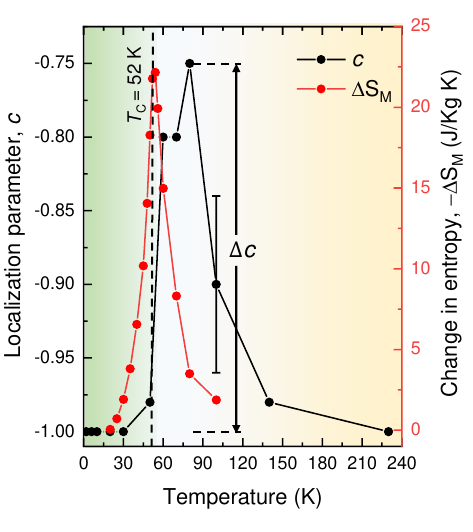}
    \vspace{-20pt}
	\caption{Temperature-dependent localization parameter $c$ extracted from the Drude-Smith (DS) modeling of the THz conductivity (black curve), showing a maximum change of $\Delta c$ close to the $T_{\rm C}$. The temperature-dependent change in the magnetic entropy (adapted with permission from Ref.~[8]) is plotted for correspondence. The yellow- and green-shaded regions show the paramagnetic and the ferromagnetic phases, respectively. The vertical dashed-line marks the $T_{\rm C}$ of the system. The error bar indicates the average standard errors obtained from the Drude-Smith modeling of the experimental data.}
	\label{figS3}
\end{figure}

Within the Drude-Smith model, the localization parameter provides an information on the degree of confinement (and hence the carrier scattering) in the system. The value of $c$ can range from -1 to 0 where $c=-1$ would correspond to maximum amount of carrier scattering in the system. By modeling the experimentally obtained THz conductivity using the Drude-Smith model, we extract the temperature dependence of the localization parameter ($c$), which is plotted in Fig.~\ref{figS3}. We find that $c$ lies within -1 to -0.75, indicating a higher degree of scattering within the system. In the high temperature region (i.e., $T > T_{\rm C}$), when the spins are randomly oriented, more carrier scattering prevails leading to $c$ being close to -1. On the other hand, when we are in the low temperature range (i.e., $T < T_{\rm C}$), the emergence of domains creates an additional degree of confinement in the system. Thus, the localization parameter again assumes a value close to $-1$. Near the phase transition, however, as the spins start to order, the material experiences a maximum change in its localization environment. Such a change in the electronic environment can potentially act as a seed to the dramatic change in the material's magnetic entropy, eventually leading to a giant magnetocaloric effect in PrSi as reported in Ref.~[8]. To bring out this correspondence, we overlay the magnetocaloric data as a function of temperature in Fig.~\ref{figS3}.

\begin{figure}[t!]
    \centering
    \includegraphics[width=0.7\columnwidth]{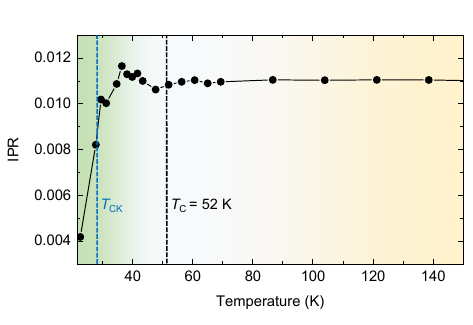}
    \vspace{-20 pt}
	\caption{Temperature-dependence inverse participation ratio (IPR). The yellow- and green-shaded regions show the paramagnetic and the ferromagnetic phases, respectively. The vertical black and blue dashed-line marks the $T_{\rm C}$ and $T_{\rm CK}$ of the system.}
	\label{figS4}
\end{figure}

\section{Inverse Participation Ratio (IPR)}
We calculated the inverse participation ratio (IPR) to quantify the localization properties of the spin channels. We used the standard definition of IPR for spin channel $\sigma$ as: $\sum_{{\bf r}_i}|\psi_{\alpha,\sigma}({\bf r}_i)|^{4}$, where $\psi_{\alpha,\sigma}({\bf r}_i)$ is the amplitude of the $\alpha^{\rm th}$ eigenstate at position ${\bf r}_i$. We choose a small window in energy around the chemical potential and take an average of the IPR over the eigenstates within that window. The resulting IPR is further averaged over 100 Monte-Carlo samples at every temperature and five independent Monte-Carlo runs with different random number seeds. In Fig.~\ref{figS4}, we show the temperature dependence of the IPR for the majority spin channel. It is known that IPR scales as inverse square of the localization length scales, which systematically allows us to quantify the increased scattering of the itinerant electrons from the localized moments that become more disordered with increasing temperature. The enhanced scattering thereby reduces the mean free path of the electrons and leads to a suppression of the Drude weight as $\omega \to 0$, consistent with literature~\cite{Kumar2005}. We have also checked that for $T << T_{\rm CK}$, the IPR reduces to $1/L^2$, $L$ being the linear system dimension, where the local moments are aligned in a ferromagnetic order and offer negligible scattering of the itinerant electrons.

\begin{figure}[b!]
    \centering
	\includegraphics[width=0.45\columnwidth]{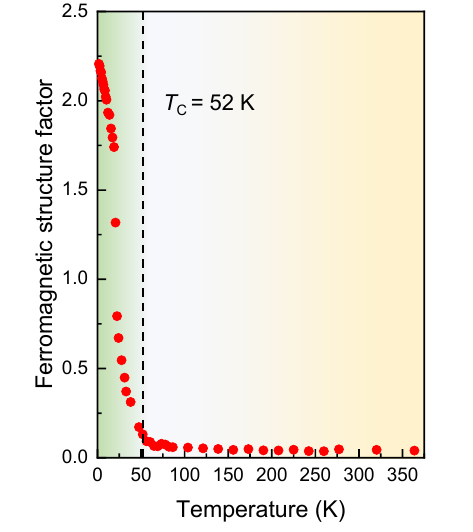}
    \vspace{-20pt}
	\caption{Temperature-dependent ferromagnetic structure factor, showing the onset of magnetic order in the system. The yellow- and green-shaded regions show the paramagnetic and the ferromagnetic phases, respectively. The vertical dashed-line marks the $T_{\rm C}$ of our system.}
	\label{figS5}
\end{figure}

\section{The structure factor}
For the characterization of the magnetic ordering in our system, we look at the magnetic structure factor $S(q)$, given by
\begin{equation*}
 S(q)=\frac{1}{N^2}\sum_{i,j}e^{i\textbf{q}\cdot(r_i-r_j)}\left<\textbf{S}_i\cdot\textbf{S}_j\right >,
\end{equation*}
where $\textbf{q}={0,0}$ is the wave-vector considered, as we are interested in the ferromagnetic order. $N$ is the dimension of the system. $\textbf{S}_i$ and $\textbf{S}_j$ are the spins at the $i-$th and $j-$th site, respectively. $r_i$ and $r_j$ give the position of the respective spins. The structure factor gives an estimation for the ordering temperature, see Fig.~\ref{figS5}.

\section{Frequency-integrated spectral weight}
\begin{figure}[b!]
    \centering
	\includegraphics[width=0.7\columnwidth]{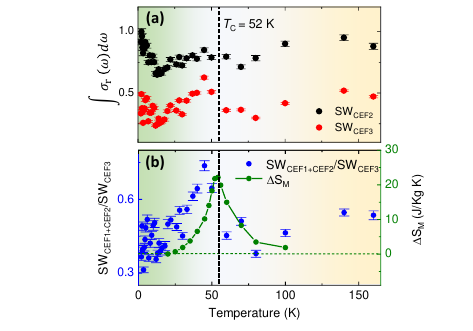}
    \vspace{-20pt}
	\caption{{\bf (a)} Temperature-dependent frequency-integrated spectral weight as per the Kubo's formula. {\bf (b)} Ratio of the spectral weights corresponding to CEF$_2$ and CEF$_3$ as a function of temperature. The temperature-dependent change in the magnetic entropy (adapted with permission from Ref.~[8]) is plotted for correspondence. The yellow- and the green-shaded regions show the paramagnetic and the ferromagnetic phases, respectively. The vertical dashed-line marks the $T_{\rm C}$ of our system. Here, the error bars are associated with the standard errors of numerical integration.}
	\label{figS6}
\end{figure}

\begin{figure}[b!]
    \centering
	\includegraphics[width=0.6\columnwidth]{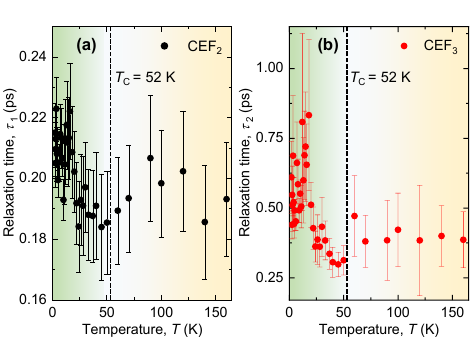}
    \vspace{-20pt}
	\caption{The temperature dependence of the relaxation times obtained from DL modeling, corresponding to {\bf (a)} CEF$_2$ (i.e., $\tau_1$) and {\bf (b)} CEF$_3$ (i.e., $\tau_2$). The yellow- and the green-shaded regions show the paramagnetic and the ferromagnetic phases, respectively. The vertical dashed-lines mark the $T_{\rm C}$ of our system.}
	\label{figS7}
\end{figure}

To verify our fitting results obtained from the double-Lorentz model, we evaluated the frequency-integrated spectral weight of the corresponding peaks. We know from Kubo's formula,
\begin{equation}
\int \sigma_r(\omega) \, d\omega = \frac{\pi n_{\text{eff}} e^2}{2 m^*}
\end{equation}
where, $\sigma_{\rm r}$ is the real part of the THz conductivity, $n_{\rm eff}$ is the effective number of electrons involved in the absorption process, $e$ is the electronic charge, and $m^*$ is the effective mass of the charge carriers~\cite{Basov2011}. An integration on the CEF peak would give us an idea on the number of carriers involved in the process. In Fig.~\ref{figS6}a, we see that the occupation corresponding to CEF$_2$ transition is always higher at all temperature. We expect that because that is the lower excited state. However, at T$_c$ the occupation corresponding to CEF$_3$ increases. The relative spectral weight plotted in Fig.~\ref{figS6}b as a function of temperature, clearly corroborates our results shown in Fig.4c of the main manuscript.

\section{CEF line-widths}
We observe that the relaxation time corresponding to both the CEF peaks increases as we lower the temperature (see Figs.~\ref{figS7}a and~\ref{figS7}b). In other words, the linewidth (1/$\tau$) corresponding to the peaks decreases, which not only supports the underlying thermal broadening in the material but also corroborates the onset of the magnetic ordering of the system. This dependence substantiates that our second approach with modeling of the CEF environment using DL oscillators throughout the temperature range. Note that we are not expecting a dramatic increase of the line-widths as increase the temperature since the system is metallic, where the CEF transitions are buried deep below the Fermi level.